\documentclass{iaus}
\usepackage{graphicx}
\usepackage{natbib}

\title[Magnetic coronae] 
{Magnetic coronae of active main-sequence stars}

\author[M. Jardine, J.-F. Donati]   
{Moira Jardine$^1$ and Jean-Francois Donati$^2$}

\affiliation{$^1$SUPA, School of Physics and Astronomy, University of St Andrews, North Haugh, St Andrews, KY16 9SS, UK \\ email: {\tt mmj@st-andrews.ac.uk} \\[\affilskip]
$^2$LATT, CNRS--UMR 5572, Obs.\ Midi-Pyr\'en\'ees, 14 Av.\ E.~Belin, F--31400 Toulouse, France \\email: {\tt donati@ast.obs-mip.fr}} 

\pubyear{2008}
\volume{259}  
\pagerange{***--***}
\jname{Cosmic magnetic fields: from planets to stars and galaxies}
\editors{Klaus G. Strassmeier, Alexander G. Kosovichev \& John Beckmann, eds.}
\begin{document}

\maketitle

\begin{abstract}
The coronal structure of main sequence stars continues to puzzle us. While the solar corona is relatively well understood, it has become clear that even stars of the same mass as the Sun can display very non-solar coronal behaviour, particularly if they are rapid rotators or in a binary system. At masses greater than and also less than that of the Sun, the non-solar internal structure appears to affect both the geometry and dynamics of the stellar corona and the nature of the X-ray and radio emission. In this talk I will describe some recent advances in our understanding of the structure of the coronae of some of the most active (and interesting) main sequence stars.
\end{abstract}

\firstsection 
\section{Introduction}

During the course of this Symposium we have already heard about the range of surface magnetic fields that can be found on different types of stars across the HR diagram. In this talk I want to address the slightly different question of the nature of the coronae of active stars on the main sequence. In the last 5 to 10 years it has become apparent that the change in internal structure with mass along the main sequence can have profound effects on the types of magnetic fields that can be generated. At the high-mass end of the main sequence, the presence of significant magnetic fields on some stars with radiative interiors challenges current dynamo theories that rely on convective processes. One possibility is that they generate fields in their convective cores, although this raises the question of how to transport the flux to the surface \citep{charbonneau_macgregor_dynamo_01,brun_dynamo_05}. They can also generate magnetic fields in the radiative zone, but a very non-solar dynamo process \citep{spruit_dynamo_02,tout_pringle_dynamo_95,macdonald_mullan_dynamo_04,mullan_macdonald_dynamo_05,maeder_meynet_dynamo_05}, Alternatively, the fields may be fossils, left over from the early stages of the formation of the star \citep{moss_review_01,braithwaite_spruit_fossilfields_04,braithwaite_nordlund_dynamo_06}. 

\begin{figure}[t]
\begin{center}

  \includegraphics[width=2.5in]{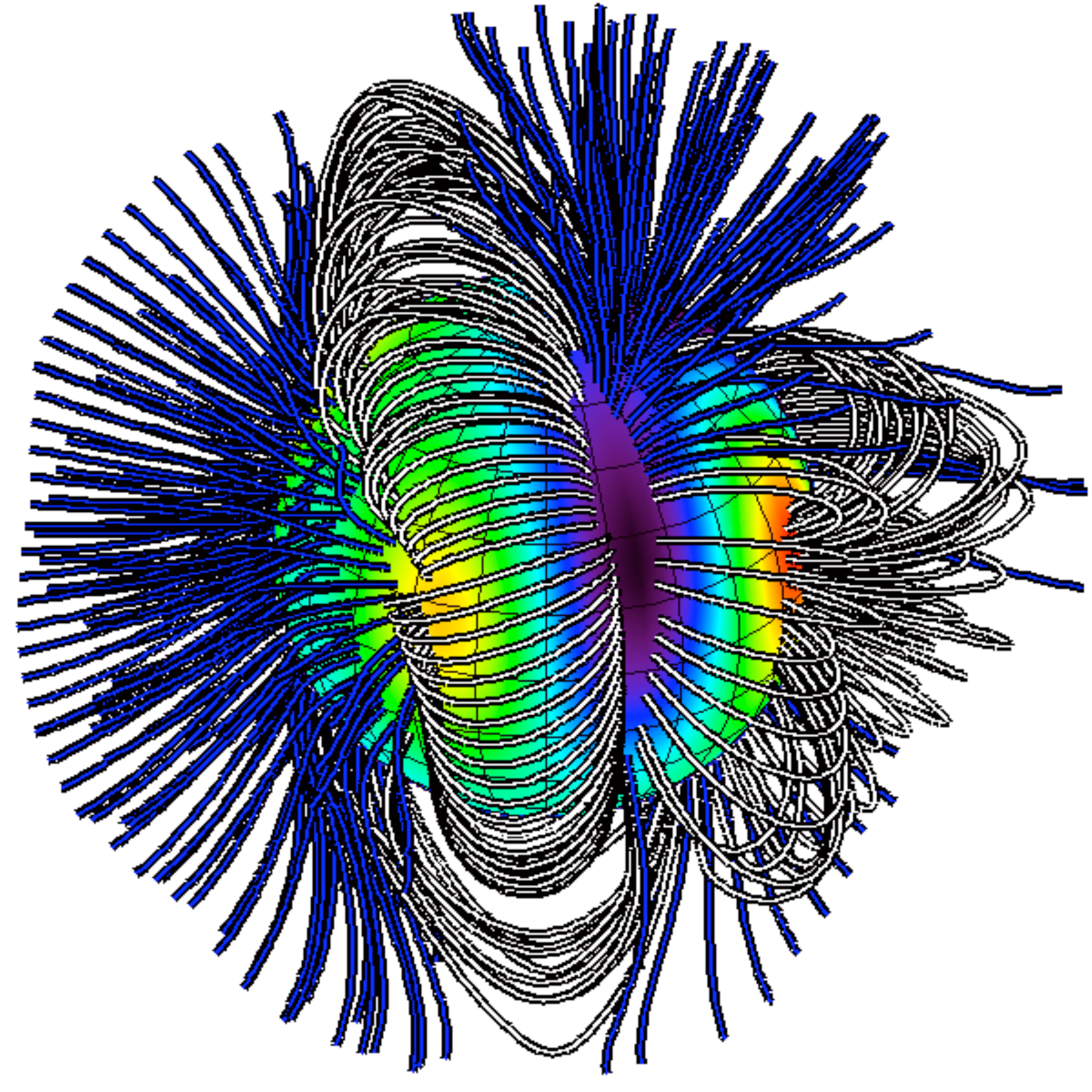} 
  \includegraphics[width=2.5in]{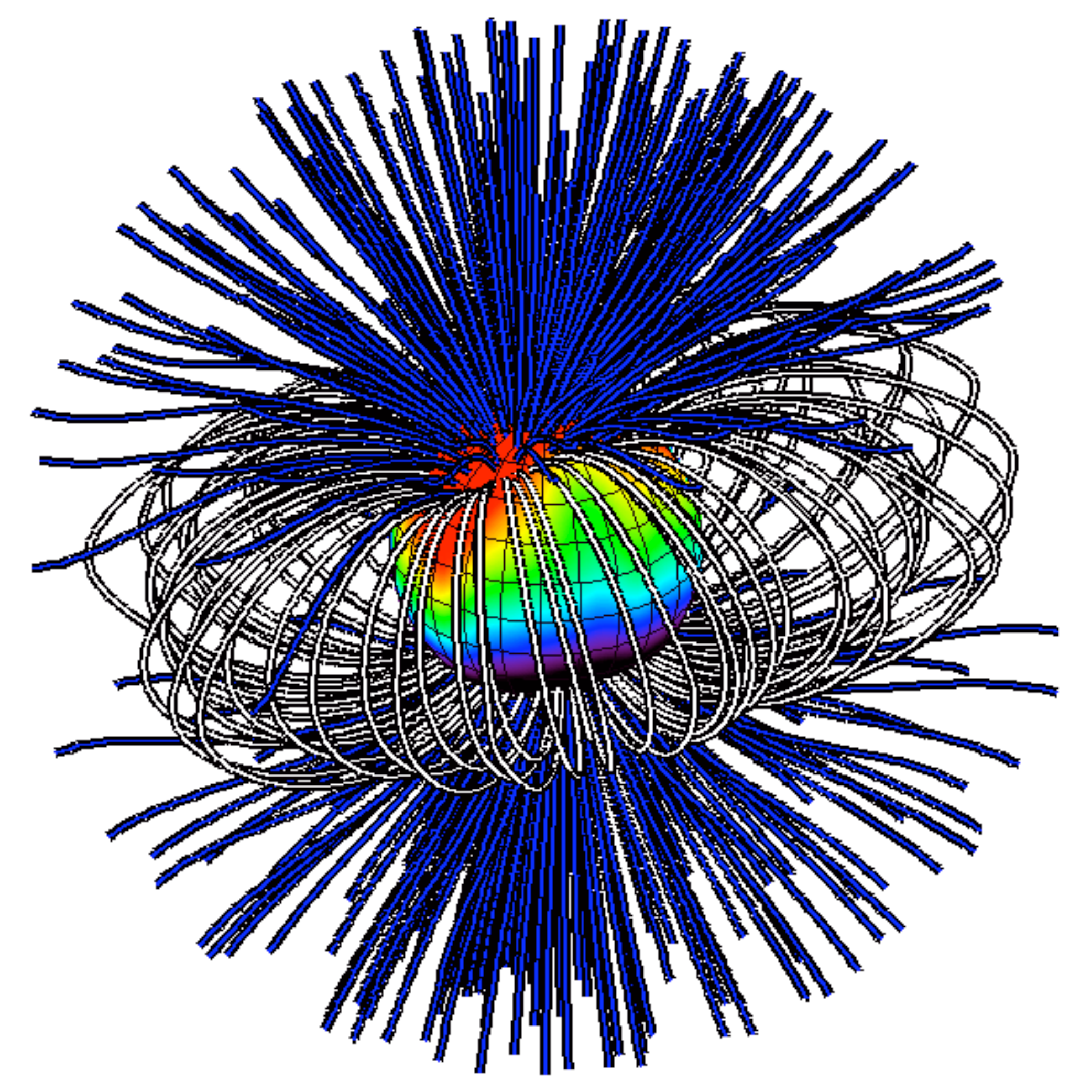} 

 \caption{Closed field lines (white) and open field lines (blue) extrapolated from a Zeeman-Doppler image of Tau Sco (left) and V374 Peg (right).}
   \label{fig1}
\end{center}
\end{figure}

The fossil field explanation is perhaps the one most likely to explain the complex field detected on Tau Sco \citep{donati06tausco}. At 15 M$_\odot$ it has a radiative interior, and yet as shown in Fig. \ref{fig1} it displays a complex, strong field. If this is a fossil field, it might be expected to be a simple dipole, but the very youth of this star, at only a million years, may be the reason why the higher-order field components have not yet decayed away. Interestingly, Tau Sco shows H$_\alpha$ absorption features that are very similar to prominence signatures in lower mass stars. These prominences are clouds of cool, mainly neutral hydrogen, confined within the million-degree gas of the stellar corona. They are observed as transient absorption features crossing the H-$\alpha$ profile when the prominence crosses in front of the stellar disk as seen by an observer \citep{cameron89cloud,cameron89eject,cameron92alpper,jeffries93,byrne96hkaqr,eibe98re1816,barnes20PZTel,donati20RXJ}. High mass stars like Tau Sco also show similar H$_\alpha$ absorption features, but in this case they are attributed to a ``wind-compressed disk'' that forms when sections of the very massive wind emanating from different parts of the stellar disk collide and cool \citep{townsend_winds_05}. 

Very low mass stars also have an internal structure that is very different from that of the Sun in that convection may extend throughout their interiors. In the absence of a tachocline, these stars cannot support a solar-like interface dynamo, yet they, like the high mass stars, exhibit observable magnetic fields. The mechanism by which they generate these magnetic fields has received a great deal of attention recently. While a decade or so ago, it was believed that these stars could only generate small-scale magnetic fields  \citep{durney_turb_dynamo_93, cattaneo_dynamo_99}, more recent studies have suggested that large scale fields may be generated. These models differ, however, in their predictions for the form of this field and the associated latitudinal differential rotation. They predict that the fields should be either axisymmetric with pronounced differential rotation \citep{dobler_dynamos_06}, non-axisymmtric with minimal differential rotation \citep{kuker_rudiger_97,kuker_rudiger_99,chabrier_kuker_06}.

In contrast, as shown in Fig. \ref{fig1} the very low mass fully-convective star V374 Peg has a very simple, dipolar field \citep{donati06v374peg}. The highly-symmetric nature of the field and the absence of a measureable differential rotation are consistent with the recent models of \citet{browning_dynamo_08}. It is unfortunately not possible at present to detect any prominences that might be present on these very low mass stars because they stars are intrinsically too faint. Their detection would, however, be a very clear test of the magnetic structure, since in a simple dipole any prominences should, by symmetry, form in the equatorial plane.

\section{How do we model stellar coronae?}
In this short talk I will not describe any further the magnetic fields of stars at the extreme ends of the main sequence. I will concentrate instead on the solar mass stars for which many more observations exist. So how do we construct models of stellar coronae? The first step is to obtain a map of the surface magnetic field. This is most commonly done using the technique of Zeeman-Doppler imaging  \citep{donati97abdor95,donati99abdor96}. These maps typically show a complex distribution of surface spots that is often very different from that of the Sun, with spots and mixed polarity flux elements extending over all latitudes up to the pole \citep{strassmeier96table}. 

\begin{figure*}[t]
\begin{center}

 \includegraphics[width=2.5in]{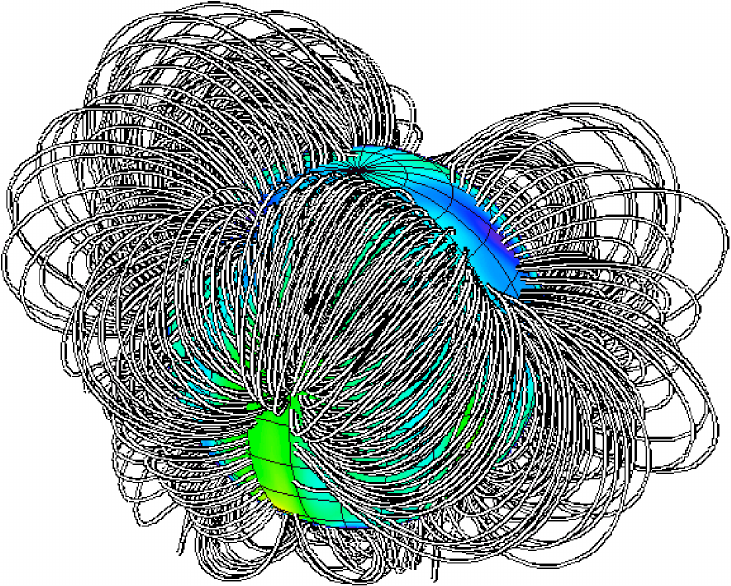} 
  \includegraphics[width=2.5in]{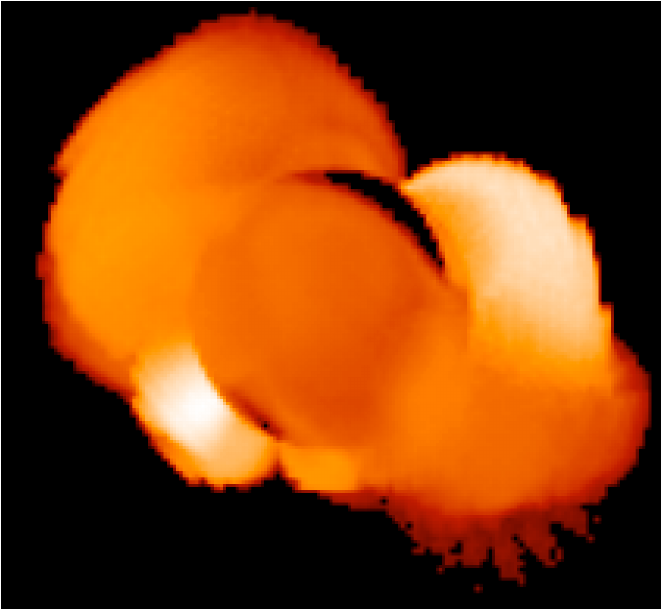} 

 \caption{Closed field lines (left) and corresponding X-ray image (right) for the rapidly-rotating star LQ Hya. A coronal temperature of 10$^6$K is assumed.}
   \label{fig2}
\end{center}
\end{figure*}

From this magnetogram we can extrapolate the coronal magnetic field using a {\it Potential Field Source Surface} method \citep{altschuler69,jardine99ccf,jardine2001eqm,jardine02structure,mcivor03polar}, or using non-potential fields \citep{donati01,hussain02nonpot}. By assuming that the gas trapped on these field lines is in isothermal, hydrostatic equilibrium, we can determine the coronal gas pressure, subject to an assumption for the gas pressure at the base of the corona. We assume that it is proportional to the magnetic pressure, i.e. $p_0 \propto B_0^2$, where the constant of proportionality is determined by comparison with X-ray emission measures \citep{jardine02xray,jardine_TTS_06,gregory_rotmod_06}. For an optically thin coronal plasma, this then allows us to produce images of the X-ray emission, as shown in Fig.\ref{fig2}.

This immediately highlights one of the greatest puzzles about active solar mass stars, which is the extent of their coronae. X-ray spectra reveal coronal densities so high that they can only be confined in a compact corona \citep{dupree93,schrijver95,brickhouse98,maggio2000,gudel01XMM,sanz_forcada_abdor_03}. Despite this, as discussed in the introduction, these stars often harbour so-called ``slingshot prominences''.  In many instances these prominences re-appear on subsequent stellar rotations, often with some change in the time taken for the absorption feature to travel through the line profile. As many as six may be present in the observable hemisphere.  What is most surprising about them is their location, which is inferred from the  time taken for the absorption features to travel through the line profile. Values of several stellar radii from the stellar rotation axis are typically found, suggesting that the confinement of these clouds is enforced out to very large distances. Indeed the preferred location of these prominences appears to be at or beyond the equatorial stellar co-rotation radius, where the inward pull of gravity is exactly balanced by the outward pull of centrifugal forces. Beyond this point, the effective gravity (including the centrifugal acceleration) points outwards and the presence of a restraining force, such as the tension in a closed magnetic loop, is required to hold the prominence in place against centrifugal ejection. The presence of these prominences therefore immediately requires that the star have many closed loop systems that extend out for many stellar radii. It is difficult to reconcile this need for an extended corona with the high densities inferred from X-ray studies.

\begin{figure*}
\begin{center}

\includegraphics[width=2.5in]{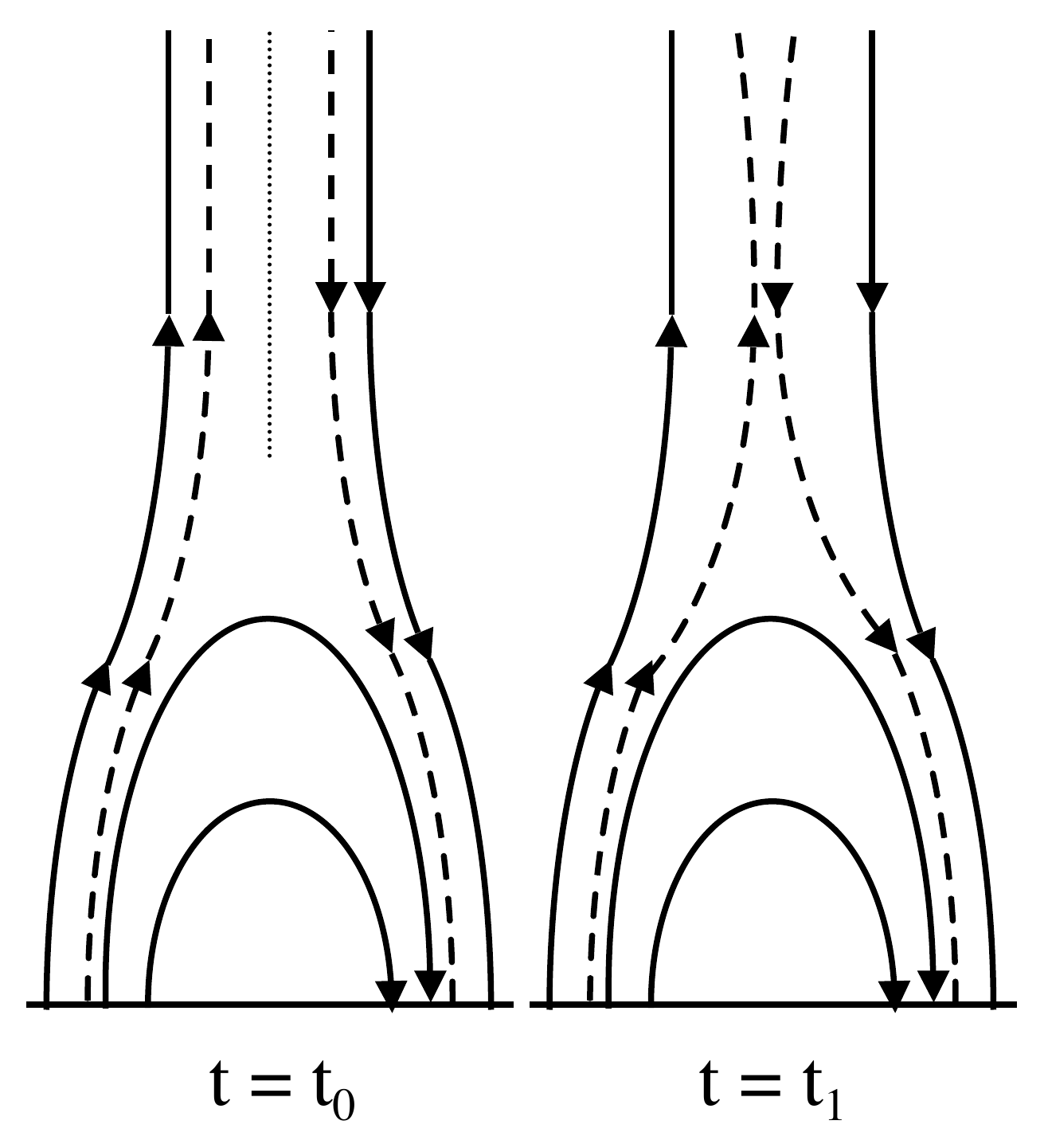}
\includegraphics[width=2.5in]{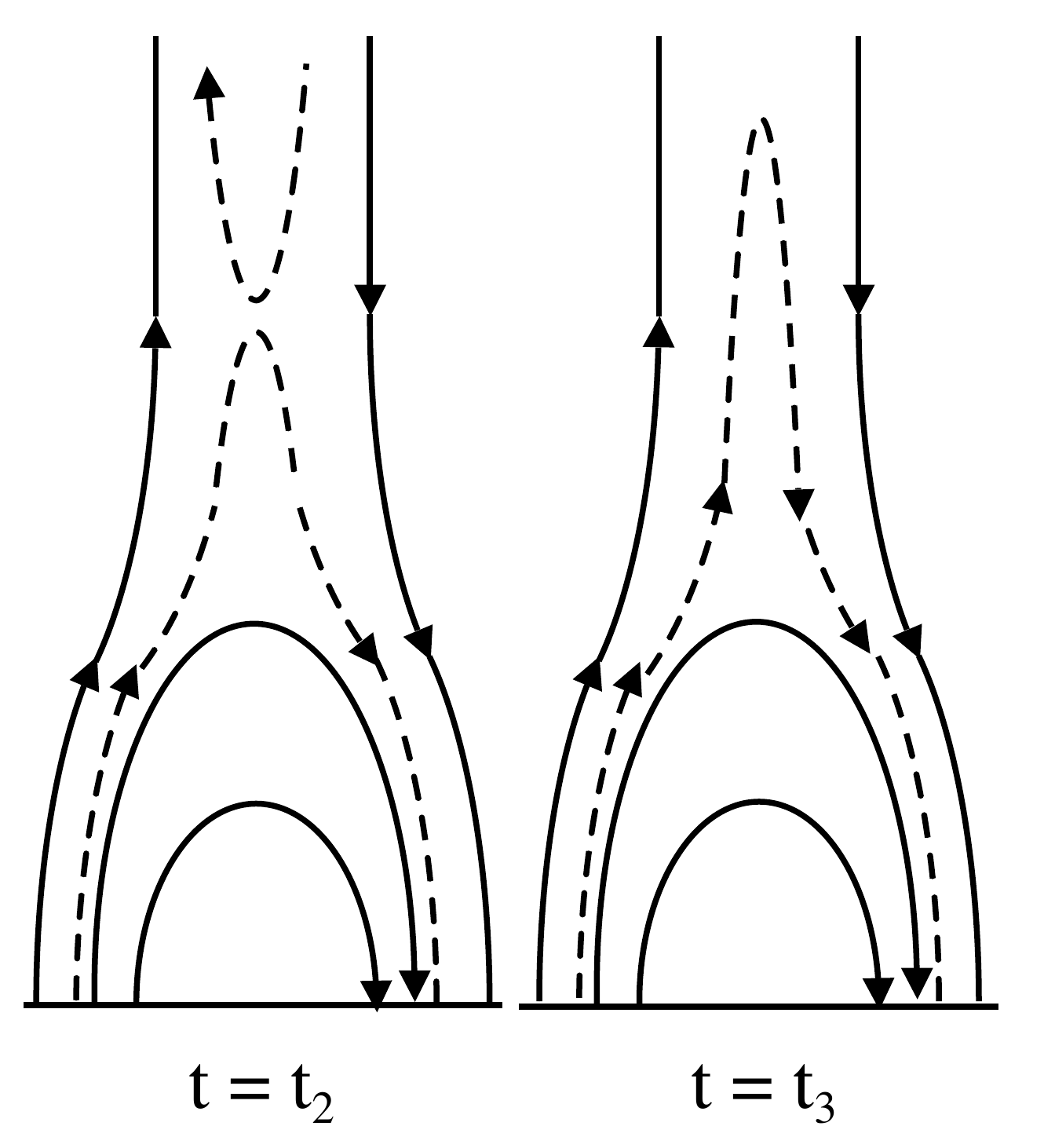}

 \caption{A schematic diagram of the formation of prominence-bearing loops. Initially, at $t=t_0$ a current sheet is present above the cusp of a helmet streamer. Reconnection at in the current sheet at $t=t_1$ produces a closed loop at $t=t_2$. The stellar wind continues to flow until pressure balance is restored, thus increasing the density in the top of this new loop. Increased radiative losses cause the loop to cool and the change in internal pressure forces it to a new equilibrium at $t=t_3$.}

  \label{fig3}
 \end{center}
\end{figure*}
One way out of this problem is to confine the prominences in the wind region {\it beyond} the closed corona. \citet{jardine05proms} have produced a model for this that predicts a maximum height $y_m$ for the prominence as a function of the co-rotation radius, $y_K$ where
\begin{eqnarray}
\frac{y_m}{R_\star} & = & \frac{1}{2}\left(
                                             -3+\sqrt{1+\frac{8GM_\star}{R_\star^3 \omega^2}}
                                                       \right) \\
                               & = & \frac{1}{2}\left(
                                             -3+\sqrt{1+8 \left[
                                                                       \frac{y_K}{R_\star} + 1
                                                                  \right]^3}
                                                       \right).
\end{eqnarray} 
Fig. \ref{fig3} shows the sequence of events that might lead to the formation of one of these ``slingshot'' prominences. The stellar wind flows along the open field lines that bound a closed field region, forming a helmet streamer. If the current sheet that forms between these oppositely-directed field lines reconnects, then a loop of magnetic field will be formed. The stellar wind will continue to flow for a short time, until pressure balance is re-established with a new field configuration. Jardine $\&$ van Ballegooijen (2006) showed that a new, cool equilibrium was possible which could reach out well beyond the co-rotation radius. The distribution of prominence heights shown in \cite{dunstone_speedymic_I_08,dunstone_speedymic_II_08} for the ultra-fast rotator Speedy Mic shows prominences forming up to (but not significantly beyond) this maximum height.

\section{Conclusions}
This new model for the support of prominences in rapid rotators may indeed resolve the conflicting pieces of evidence about the extent of the stellar corona, allowing for a compact X-ray corona {\it and} the support of many prominences at large distance from the stellar rotation axis. More observations are clearly needed, however, to test this theory. The nature of the large-scale field is difficult to determine because it may well be relatively dark in X-ray emission. This is the case on the Sun, where the X-ray corona is compact, but the white-light corona (as seen in eclipses) is much more extended. On very young stars that are still accreting, we have the opportunity to observe the large-scale field since it is ``illuminated'' by the accretion process. On such young stars, it seems that the large-scale field is very simple, even when the X-ray corona is highly structured \citep{donati_v2129oph_07,donati_bptau_08,jardine_v2129oph_08,gregory_bptau_v2129oph_08}. The degree of complexity of stellar coronae and the nature of their large-scale field may therefore be a function of their evolutionary state,  as well as with their mass when on the main sequence. The change in internal structure that takes place when young stars develop a radiative core may be reflected in the different geometries of the magnetic fields that they generate and hence the structures of their coronae.  Low mass stars, of course, remain fully convective and so may retain the simple field structures that they had in their youth even through their main-sequence lifetimes. This may have consequences not only for their X-ray emission, but also for their rotational histories, since their magnetic fields are crucial to their ability to spin down from the pre-main sequence phase. Over the next few years, as information on distributions of stellar rotation rates becomes available through CoRoT, this will undoubtedly prove to be a fruitful line of research.


\begin{thebibliography}{54}
\expandafter\ifx\csname natexlab\endcsname\relax\def\natexlab#1{#1}\fi

\bibitem[{{Altschuler} \& {Newkirk, Jr.}(1969)}]{altschuler69}
{Altschuler}, M.~D. \& {Newkirk, Jr.}, G. 1969, Solar~Phys., 9, 131

\bibitem[{Barnes {et~al.}(2000)Barnes, Collier~Cameron, James, \&
  Donati}]{barnes20PZTel}
Barnes, J., Collier~Cameron, A., James, D.~J., \& Donati, J.-F. 2000, MNRAS,
  314, 162

\bibitem[{{Braithwaite} \& {Nordlund}(2006)}]{braithwaite_nordlund_dynamo_06}
{Braithwaite}, J. \& {Nordlund}, A. 2006, A\&A, 450, 1077

\bibitem[{{Braithwaite} \& {Spruit}(2004)}]{braithwaite_spruit_fossilfields_04}
{Braithwaite}, J. \& {Spruit}, H.~C. 2004, Nature, 431, 819

\bibitem[{{Brickhouse} \& {Dupree}(1998)}]{brickhouse98}
{Brickhouse}, N. \& {Dupree}, A. 1998, ApJ, 502, 918

\bibitem[{{Browning}(2008)}]{browning_dynamo_08}
{Browning}, M. 2008, ApJ, in press

\bibitem[{{Brun} {et~al.}(2005){Brun}, {Browning}, \&
  {Toomre}}]{brun_dynamo_05}
{Brun}, A.~S., {Browning}, M.~K., \& {Toomre}, J. 2005, ApJ, 629, 461

\bibitem[{{Byrne} {et~al.}(1996){Byrne}, {Eibe}, \& {Rolleston}}]{byrne96hkaqr}
{Byrne}, P., {Eibe}, M., \& {Rolleston}, W. 1996, A\&A, 311, 651

\bibitem[{{Cattaneo}(1999)}]{cattaneo_dynamo_99}
{Cattaneo}, F. 1999, ApJ, 515, L39

\bibitem[{{Chabrier} \& {K\"uker}(2006)}]{chabrier_kuker_06}
{Chabrier}, G. \& {K\"uker}, M. 2006, A\&A, 446, 1027

\bibitem[{{Charbonneau} \& {MacGregor}(2001)}]{charbonneau_macgregor_dynamo_01}
{Charbonneau}, P. \& {MacGregor}, K.~B. 2001, ApJ, 559, 1094

\bibitem[{Collier~Cameron \& Robinson(1989{\natexlab{a}})}]{cameron89eject}
Collier~Cameron, A. \& Robinson, R.~D. 1989{\natexlab{a}}, MNRAS, 238, 657

\bibitem[{Collier~Cameron \& Robinson(1989{\natexlab{b}})}]{cameron89cloud}
Collier~Cameron, A. \& Robinson, R.~D. 1989{\natexlab{b}}, MNRAS, 236, 57

\bibitem[{Collier~Cameron \& Woods(1992)}]{cameron92alpper}
Collier~Cameron, A. \& Woods, J.~A. 1992, MNRAS, 258, 360

\bibitem[{{Dobler} {et~al.}(2006){Dobler}, {Stix}, \&
  {Brandenburg}}]{dobler_dynamos_06}
{Dobler}, W., {Stix}, M., \& {Brandenburg}, A. 2006, ApJ, 638, 336

\bibitem[{{Donati}(2001)}]{donati01}
{Donati}, J.-F. 2001, LNP Vol.~573: Astrotomography, Indirect Imaging Methods
  in Observational Astronomy, 573, 207

\bibitem[{Donati \& Collier~Cameron(1997)}]{donati97abdor95}
Donati, J.-F. \& Collier~Cameron, A. 1997, MNRAS, 291, 1

\bibitem[{Donati {et~al.}(1999)Donati, Collier~Cameron, Hussain, \&
  Semel}]{donati99abdor96}
Donati, J.-F., Collier~Cameron, A., Hussain, G., \& Semel, M. 1999, MNRAS, 302,
  437

\bibitem[{{Donati} {et~al.}(2006{\natexlab{a}}){Donati}, {Forveille},
  {Cameron}, {Barnes}, {Delfosse}, {Jardine}, \& {Valenti}}]{donati06v374peg}
{Donati}, J.-F., {Forveille}, T., {Cameron}, A.~C., {et~al.}
  2006{\natexlab{a}}, Science, 311, 633

\bibitem[{{Donati} {et~al.}(2006{\natexlab{b}}){Donati}, {Howarth}, {Jardine},
  {Petit}, {Catala}, {Landstreet}, {Bouret}, {Alecian}, {Barnes}, {Forveille},
  {Paletou}, \& {Manset}}]{donati06tausco}
{Donati}, J.-F., {Howarth}, I.~D., {Jardine}, M.~M., {et~al.}
  2006{\natexlab{b}}, MNRAS, 370, 629

\bibitem[{{Donati} {et~al.}(2007){Donati}, {Jardine}, {Gregory}, {Petit},
  {Bouvier}, {Dougados}, {M{\'e}nard}, {Cameron}, {Harries}, {Jeffers}, \&
  {Paletou}}]{donati_v2129oph_07}
{Donati}, J.-F., {Jardine}, M.~M., {Gregory}, S.~G., {et~al.} 2007, MNRAS, 380,
  1297

\bibitem[{{Donati} {et~al.}(2008){Donati}, {Jardine}, {Gregory}, {Petit},
  {Paletou}, {Bouvier}, {Dougados}, {M{\'e}nard}, {Cameron}, {Harries},
  {Hussain}, {Unruh}, {Morin}, {Marsden}, {Manset}, {Auri{\`e}re}, {Catala}, \&
  {Alecian}}]{donati_bptau_08}
{Donati}, J.-F., {Jardine}, M.~M., {Gregory}, S.~G., {et~al.} 2008, MNRAS, 386,
  1234

\bibitem[{Donati {et~al.}(2000)Donati, Mengel, Carter, Cameron, \&
  Wichmann}]{donati20RXJ}
Donati, J.-F., Mengel, M., Carter, B., Cameron, A., \& Wichmann, R. 2000,
  MNRAS, 316, 699

\bibitem[{{Dunstone} {et~al.}(2008{\natexlab{a}}){Dunstone}, {Hussain},
  {Cameron}, {Marsden}, {Jardine}, {Barnes}, {Ramirez Velez}, \&
  {Donati}}]{dunstone_speedymic_I_08}
{Dunstone}, N.~J., {Hussain}, G.~A.~J., {Cameron}, A.~C., {et~al.}
  2008{\natexlab{a}}, MNRAS, 387, 1525

\bibitem[{{Dunstone} {et~al.}(2008{\natexlab{b}}){Dunstone}, {Hussain},
  {Collier Cameron}, {Marsden}, {Jardine}, {Stempels}, {Ramirez Velez}, \&
  {Donati}}]{dunstone_speedymic_II_08}
{Dunstone}, N.~J., {Hussain}, G.~A.~J., {Collier Cameron}, A., {et~al.}
  2008{\natexlab{b}}, MNRAS, 387, 481

\bibitem[{{Dupree} {et~al.}(1993){Dupree}, {Brickhouse}, {Doschek}, {Green}, \&
  {Raymond}}]{dupree93}
{Dupree}, A., {Brickhouse}, N., {Doschek}, G., {Green}, J., \& {Raymond}, J.
  1993, ApJ, 418, L41

\bibitem[{{Durney} {et~al.}(1993){Durney}, {De Young}, \&
  {Roxburgh}}]{durney_turb_dynamo_93}
{Durney}, B.~R., {De Young}, D.~S., \& {Roxburgh}, I.~W. 1993, Solar~Phys.,
  145, 207

\bibitem[{{Eibe}(1998)}]{eibe98re1816}
{Eibe}, M.~T. 1998, A\&A, 337, 757

\bibitem[{{Gregory} {et~al.}(2006){Gregory}, {Jardine}, {Cameron}, \&
  {Donati}}]{gregory_rotmod_06}
{Gregory}, S.~G., {Jardine}, M., {Cameron}, A.~C., \& {Donati}, J.-F. 2006,
  MNRAS, 373, 827

\bibitem[{{Gregory} {et~al.}(2008){Gregory}, {Matt}, {Donati}, \&
  {Jardine}}]{gregory_bptau_v2129oph_08}
{Gregory}, S.~G., {Matt}, S.~P., {Donati}, J.-F., \& {Jardine}, M. 2008, MNRAS,
  in press

\bibitem[{{G\"udel} {et~al.}(2001){G\"udel}, {Audard}, {Briggs}, {Haberl},
  {Magee}, {Maggio}, {Mewe}, {Pallavicini}, \& {Pye}}]{gudel01XMM}
{G\"udel}, M., {Audard}, M., {Briggs}, K., {et~al.} 2001, A\&A, 365, L336

\bibitem[{{Hussain} {et~al.}(2002){Hussain}, {van Ballegooijen}, {Jardine}, \&
  {Collier Cameron}}]{hussain02nonpot}
{Hussain}, G.~A.~J., {van Ballegooijen}, A.~A., {Jardine}, M., \& {Collier
  Cameron}, A. 2002, ApJ, 575, 1078

\bibitem[{{Jardine} {et~al.}(1999){Jardine}, {Barnes}, {Donati}, \& {Collier
  Cameron}}]{jardine99ccf}
{Jardine}, M., {Barnes}, J., {Donati}, J.-F., \& {Collier Cameron}, A. 1999,
  MNRAS, 305, L35

\bibitem[{{Jardine} {et~al.}(2002{\natexlab{a}}){Jardine}, {Collier Cameron},
  \& {Donati}}]{jardine02structure}
{Jardine}, M., {Collier Cameron}, A., \& {Donati}, J.-F. 2002{\natexlab{a}},
  MNRAS, 333, 339

\bibitem[{{Jardine} {et~al.}(2006){Jardine}, {Collier Cameron}, {Donati},
  {Gregory}, \& {Wood}}]{jardine_TTS_06}
{Jardine}, M., {Collier Cameron}, A., {Donati}, J.-F., {Gregory}, S.~G., \&
  {Wood}, K. 2006, MNRAS, 367, 917

\bibitem[{{Jardine} {et~al.}(2001){Jardine}, {Collier Cameron}, {Donati}, \&
  {Pointer}}]{jardine2001eqm}
{Jardine}, M., {Collier Cameron}, A., {Donati}, J.-F., \& {Pointer}, G. 2001,
  MNRAS, 324, 201

\bibitem[{{Jardine} {et~al.}(2008){Jardine}, {Gregory}, \&
  {Donati}}]{jardine_v2129oph_08}
{Jardine}, M., {Gregory}, S.~G., \& {Donati}, J.-F. 2008, MNRAS, in press

\bibitem[{{Jardine} \& {van Ballegooijen}(2005)}]{jardine05proms}
{Jardine}, M. \& {van Ballegooijen}, A.~A. 2005, MNRAS, 361, 1173

\bibitem[{{Jardine} {et~al.}(2002{\natexlab{b}}){Jardine}, {Wood}, {Collier
  Cameron}, {Donati}, \& {Mackay}}]{jardine02xray}
{Jardine}, M., {Wood}, K., {Collier Cameron}, A., {Donati}, J.-F., \& {Mackay},
  D.~H. 2002{\natexlab{b}}, MNRAS, 336, 1364

\bibitem[{{Jeffries}(1993)}]{jeffries93}
{Jeffries}, R. 1993, MNRAS, 262, 369

\bibitem[{{K\"uker} \& {R\"udiger}(1997)}]{kuker_rudiger_97}
{K\"uker}, M. \& {R\"udiger}, G. 1997, A\&A, 328, 253

\bibitem[{{K\"uker} \& {R\"udiger}(1999)}]{kuker_rudiger_99}
{K\"uker}, M. \& {R\"udiger}, G. 1999, in ASP Conf. Ser. 178: Workshop on
  stellar dynamos, Vol. 178, 87--96

\bibitem[{{MacDonald} \& {Mullan}(2004)}]{macdonald_mullan_dynamo_04}
{MacDonald}, J. \& {Mullan}, D.~J. 2004, MNRAS, 348, 702

\bibitem[{{Maeder} \& {Meynet}(2005)}]{maeder_meynet_dynamo_05}
{Maeder}, A. \& {Meynet}, G. 2005, A\&A, 440, 1041

\bibitem[{{Maggio} {et~al.}(2000){Maggio}, {Pallavicini}, {Reale}, \&
  {Tagliaferri}}]{maggio2000}
{Maggio}, A., {Pallavicini}, R., {Reale}, F., \& {Tagliaferri}, G. 2000, A\&A,
  356, 627

\bibitem[{{McIvor} {et~al.}(2003){McIvor}, {Jardine}, {Cameron}, {Wood}, \&
  {Donati}}]{mcivor03polar}
{McIvor}, T., {Jardine}, M., {Cameron}, A.~C., {Wood}, K., \& {Donati}, J.-F.
  2003, MNRAS, 345, 601

\bibitem[{{Moss}(2001)}]{moss_review_01}
{Moss}, D. 2001, in ASP Conference Series, Vol. 248, Magnetic fields across the
  Hertzsprung-Russell diagram, ed. S.~G.~Mathys \& D.~Wickramasinghe (San
  Francisco), 305

\bibitem[{{Mullan} \& {MacDonald}(2005)}]{mullan_macdonald_dynamo_05}
{Mullan}, D.~J. \& {MacDonald}, J. 2005, MNRAS, 356, 1139

\bibitem[{{Sanz-Forcada} {et~al.}(2003){Sanz-Forcada}, {Maggio}, \&
  {Micela}}]{sanz_forcada_abdor_03}
{Sanz-Forcada}, J., {Maggio}, A., \& {Micela}, G. 2003, A\&A, 408, 1087

\bibitem[{{Schrijver} {et~al.}(1995){Schrijver}, {Mewe}, {van den Oord}, \&
  {Kaastra}}]{schrijver95}
{Schrijver}, C., {Mewe}, R., {van den Oord}, G., \& {Kaastra}, J. 1995, A\&A,
  302, 438

\bibitem[{{Spruit}(2002)}]{spruit_dynamo_02}
{Spruit}, H. 2002, A\&A, 381, 923

\bibitem[{{Strassmeier}(1996)}]{strassmeier96table}
{Strassmeier}, K. 1996, in IAU Symposium 176: Stellar Surface Structure, ed.
  {Strassmeier, K.G.} \& {Linsky, J.L.} (Kluwer), 289--298

\bibitem[{{Tout} \& {Pringle}(1995)}]{tout_pringle_dynamo_95}
{Tout}, C.~A. \& {Pringle}, J.~E. 1995, MNRAS, 272, 528

\bibitem[{{Townsend} \& {Owoki}(2005)}]{townsend_winds_05}
{Townsend}, R.~H.~D. \& {Owoki}, S.~P. 2005, MNRAS, 357, 251

\end{thebibliography}




\end{document}